\documentclass[draft]{agujournal2019}
\usepackage{url} 
\usepackage{lineno}
\usepackage[inline]{trackchanges} 
\usepackage{soul}
\usepackage{ragged2e}
\draftfalse

\journalname{Geophysical Research Letters}

\begin{document}

%
%


\title{Machine Learning Interpretability of Outer Radiation Belt Enhancement \& Depletion Events}

%
%




\authors{Donglai Ma\affil{1}, Jacob Bortnik\affil{1}, Qianli Ma\affil{1,2}, Man Hua\affil{1}, Xiangning Chu\affil{3}}


\affiliation{1}{Department of Atmospheric and Oceanic Sciences, University of California, Los Angeles, CA, USA}
\affiliation{2}{Center for Space Physics, Boston University, Boston, MA, USA}
\affiliation{3}{Laboratory for Atmospheric and Space Physics, University of Colorado Boulder, Boulder, CO, USA}






\correspondingauthor{Donglai Ma}{dma96@atmos.ucla.edu}




\begin{keypoints}
\item We use a machine learning feature attribution method to identify key drivers in radiation belt enhancement and depletion events. 
\item The electron flux depletion, loss, or enhancement is driven by the competition between solar wind Psw and cumulative strength of substorms.
\item The average AL index following the pressure maximum has a significant correlation with the resulting flux level.
\end{keypoints}

%
%

%
%


\begin{abstract}
\justifying
We investigate the response of outer radiation belt electron fluxes to different solar wind and geomagnetic indices using an interpretable machine learning method. We reconstruct the electron flux variation during 19 enhancement and 7 depletion events and demonstrate a feature attribution analysis on the superposed epoch results for the first time. We find that the intensity and duration of the substorm sequence following an initial dropout determine the overall enhancement or depletion of electron fluxes, while the solar wind pressure drives the initial dropout in both types of events. Further statistical results from a dataset with 71 events confirm this and show a significant correlation between the resulting flux levels and the average AL index, indicating that the observed "depletion" event can be more accurately described as a "non-enhancement" event.  Our novel SHAP-Enhanced Superposed Epoch Analysis (SHESEA) method can be used as an insight discovery tool in various physical systems.
\end{abstract}

\section*{Plain Language Summary}
\justifying
This study examines the responses of relativistic electrons in Earth's radiation belt to various solar wind and geomagnetic disturbances, identifying key influencing factors. We first adopt an explainable machine learning method to understand the importance of different features during 19 enhancement and 7 depletion events. Our results directly reveal that an increase in solar wind dynamic pressure contributes to a sudden decrease in electron fluxes. Additionally, we find that the strength and duration of subsequent substorms determine whether the electron flux increases or decreases. Guided by the importance of these features as determined by our machine learning model, we carry out a statistical analysis, showing a significant correlation between the flux level and the average AL index. Our method offers advantages over traditional superposed epoch analysis since it directly shows the determining factors.

%
%

%


%
%
%
%

\section{Introduction}
The radiation belt consists of trapped energetic electrons ranging from tens of keV to several MeV \cite{baker2018space}. This hazardous radiation environment, known to cause spacecraft anomalies such as surface charging and deep dielectric discharges, is associated with enhanced fluxes of hot ($5$ -- $100$ keV) and relativistic ($>500$ keV) electrons \cite{choi2011analysis,baker2016resource}. Particularly during periods of solar wind disturbances and resulting geomagnetic activity, the outer radiation belt exhibits significant variability. Electrons can be accelerated through inward radial diffusion \cite<e.g.,>{brautigam2000radial,schulz2012particle} and local acceleration driven by magnetospheric waves \cite<e.g.,>{thorne2013rapid,ma2018quantitative}. Simultaneously, losses in the outer radiation belt primarily arise from magnetopause shadowing with enhanced outward transport, and pitch angle scattering loss \cite<e.g.,>{bortnik2006observation,onsager2002radiation}. The balance between these acceleration and loss mechanisms results in either an enhanced or depleted energetic electron population in the outer radiation belt following disturbed solar wind activity \cite<e.g.,>{reeves2003acceleration, anderson2015acceleration}.

A number of previous studies have demonstrated that geomagnetic storms can result in either an increase or a decrease of the fluxes of relativistic electrons in the outer radiation belt. \citeA{reeves2003acceleration} examined 276 storms from 1989 to 2000 revealing that approximately half of the geomagnetic storms resulted in increased radiation belt electron fluxes, roughly a quarter resulted in decreased fluxes and a quarter were roughly unchanged. \citeA{turner2013storm} analyzed the electron phase space density (PSD) of 53 storms with main phase minimum Dst $< $ $-40$ nT, revealing that of the 58\% storms resulted in relativistic electron PSD enhancement, 17\% resulted in depletion, and the rest remained unchanged. These studies primarily focused on geomagnetic storm times, identified with a threshold set by the SYM-H index. \citeA{katsavrias2019statistics} employed a superposed epoch analysis to investigate flux changes during geospace disturbances by combining both storm and non-storm events. 

Despite significantly advancing our understanding of radiation belt dynamics, these studies contain certain limitations. Particularly, these studies have faced challenges in delineating the relative importance of different driving factors that occur simultaneously. Their findings are primarily based on statistical results only and cannot reproduce the processes observed in the satellite data with the identified drivers. The traditional approach to reconstructing radiation belt dynamics uses the Fokker-Planck (FP) simulations \cite<e.g.,>{ma2016characteristic,thorne2013rapid}. However, incorporating the different and competing drivers into the FP model for a large number of storms is computationally expensive, and the simulation accuracy depends on the availability of global wave distributions and simulation boundary conditions. The temporal and spatial distribution of different waves cannot be well modeled by one of the geomagnetic indices and there is no guarantee that such parameterizations are unique.

Our previous work presents an alternative approach to radiation belt modeling using a set of neural network models to reproduce the radiation belt fluxes \cite{ma2022modeling,chu2021relativistic}. The model is driven by geomagnetic indices and solar wind parameters and successfully captures electron dynamics over long and short timescales for different energies. Building on this, \citeA{ma2023opening} further examined what’s inside the ‘Black Box’ of the machine learning (ML) model with a state-of-art ML model explanation method called SHAP (Shapley Additive exPlanations,\cite{lundberg2017unified}). The proposed framework demonstrates that the feature importance of different input features deduced from the ML model aligns well with our existing physical understanding of both storm time and non-storm time enhancement events. In this letter, we employ the same interpretable ML method in conjunction with a superposed epoch analysis to investigate the enhancement and depletion events of relativistic electrons in the radiation belt. Section 2 introduces the methodology, Section 3 presents the results and our conclusions are discussed and summarized in Section 4.
\section{Data and Methods}

In this study, we utilize an ML model that is trained on relativistic electron flux data, specifically the 909 keV channel, from the 
 Magnetic Electron Ion Spectrometer (MagEIS) instrument \cite{blake2014magnetic} aboard the Van Allen Probes \cite{mauk2014science}. This model has demonstrated remarkable accuracy when tested with out-of-sample data, giving $R^2 \sim$ 0.78 -- 0.92 \cite{ma2022modeling}. In the following study, \citeA{ma2023opening} implemented the SHAP method to provide insight into the workings of the ML model using 2-hour average values of AL, SYM-H, Psw, and solar wind speed Vsw as inputs. For each input data point $\vec{x}$ to be explained, the sum of the SHAP values corresponds to the difference between the model prediction and the average prediction of the model for only background samples:$\sum \phi_i=f(\vec{x})-E(f(\vec{x}))$ where $\phi_i$ is the SHAP value of the $i$ th feature. Positive (Negative) SHAP value $\phi_i$ indicates that the feature has a positive (negative) impact on the output value. 
 
\begin{figure}
\noindent\includegraphics[width=\textwidth]{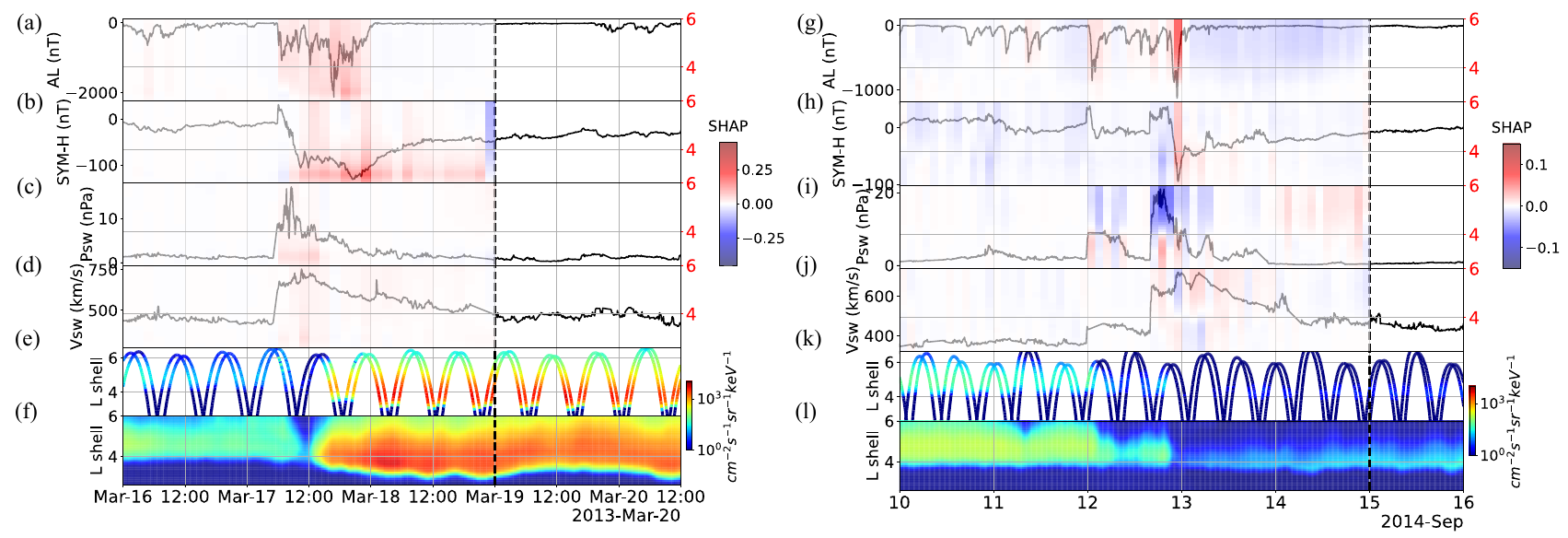}%
\caption{\label{Figure 1} Van Allen Probes observation, machine learning model and SHAP results of enhancement and depletion events. Left column: Enhancement event of 17 March 2013, input time series of the ML model:(a) AL index, (b) SYM-H index, (c) solar wind dynamic pressure, Psw, (d) solar wind speed, Vsw, (a-d) together with color-coded SHAP feature contributions for the model output at time 00 UT on 19 March 2013. (e) Observed 909 keV electron fluxes as a function of time and L-shell; (f) model reconstruction of 909 keV electron fluxes on the equatorial plane. Right column: same as (a-f) for the depletion event of 13 September 2014. The SHAP results are input feature contributions for output at time 00 UT on 15 September 2014, across all L shells. }%
\end{figure}

 Figure 1 shows examples of a typical acceleration (left panels) and depletion (right panels) event, with color-coded SHAP values superimposed on each of the corresponding inputs in Figures 1a-d and 1g-j.  In each of these panels, the y-axis on the right side of the panel indicates the particular L-shell at which the output fluxes are being affected by the current feature, in accordance with the SHAP value displayed in color.  In each event, the times at which the output fluxes to be explained are indicated by the vertical dashed lines. For the enhancement event, it is seen that the cluster of AL peaks occurring at $\sim$ 0600 --2300 UT on 17 March 2013 dominantly contributes to the acceleration of fluxes across a broad range of L-shells from $\sim$ 3--6, while the main phase and minimum of SYM-H contribute to the enhancement at lower L-shells around 3.5. For the depletion event shown in the right column, although the strong peaks in the AL index occurring before 13 September 2014 show positive SHAP values (i.e., causing flux enhancements at $\mathrm{L} > 4$), the following low intensity, continuous AL indices after 13 September 2014 and pressure enhancement occurring around 1500UT on 12 September 2014 show negative contributions to the flux at high L shells ($\mathrm{L} > 4$) which ultimately lead to a depletion of the fluxes after 13 September 2014.

 The events selected in \citeA{katsavrias2019statistics} consist of 71 intervals during the RBSP era which satisfied a set of specific conditions from at least 12 hours before each event starts: the average solar wind speed must be below 400 km/s, and pressure must be under 3 nPa; the geomagnetic  SYM-H index must be consistently over -20 nT, the AL index above -300 nT, and Bz between -5 and 5 nT. The end time of the event is taken to be the time when all parameters revert to their pre-event levels. The authors then select 20 enhancement and 8 depletion events from these 71 events for their superposed epoch analysis method. The events we use in the present study are the same as the \citeA{katsavrias2019statistics} dataset described above, except for 1 enhancement and 1 depletion event in December 2013 where the gap in solar wind data was too large for us to reasonably interpolate in order to reproduce the required ML output for our SHAP method. Another slight difference is that we define the $t_0$ epoch as the time of maximum dynamic pressure of the solar wind instead of the time of the maximum compression of the magnetopause. In addition to the standard superposed epoch analysis that was used in the above study, here we calculate the SHAP value of each feature for each event, for the output fluxes at three different times: $t_0 + 5h$, $t_0 + 1d$ and $t_0 + 2d$, in order to gain insight into how each feature controls the output. The “enhanced” superposed epoch analysis is colored with the median SHAP values of the input parameters for all the events, in order to identify the ML interpretation results as a function of storm phase, L-shell, and event type. We also calculate the ML output flux of each event and show the median flux results.

\section{Results}
 
\begin{figure}
\noindent\includegraphics[width=\textwidth]{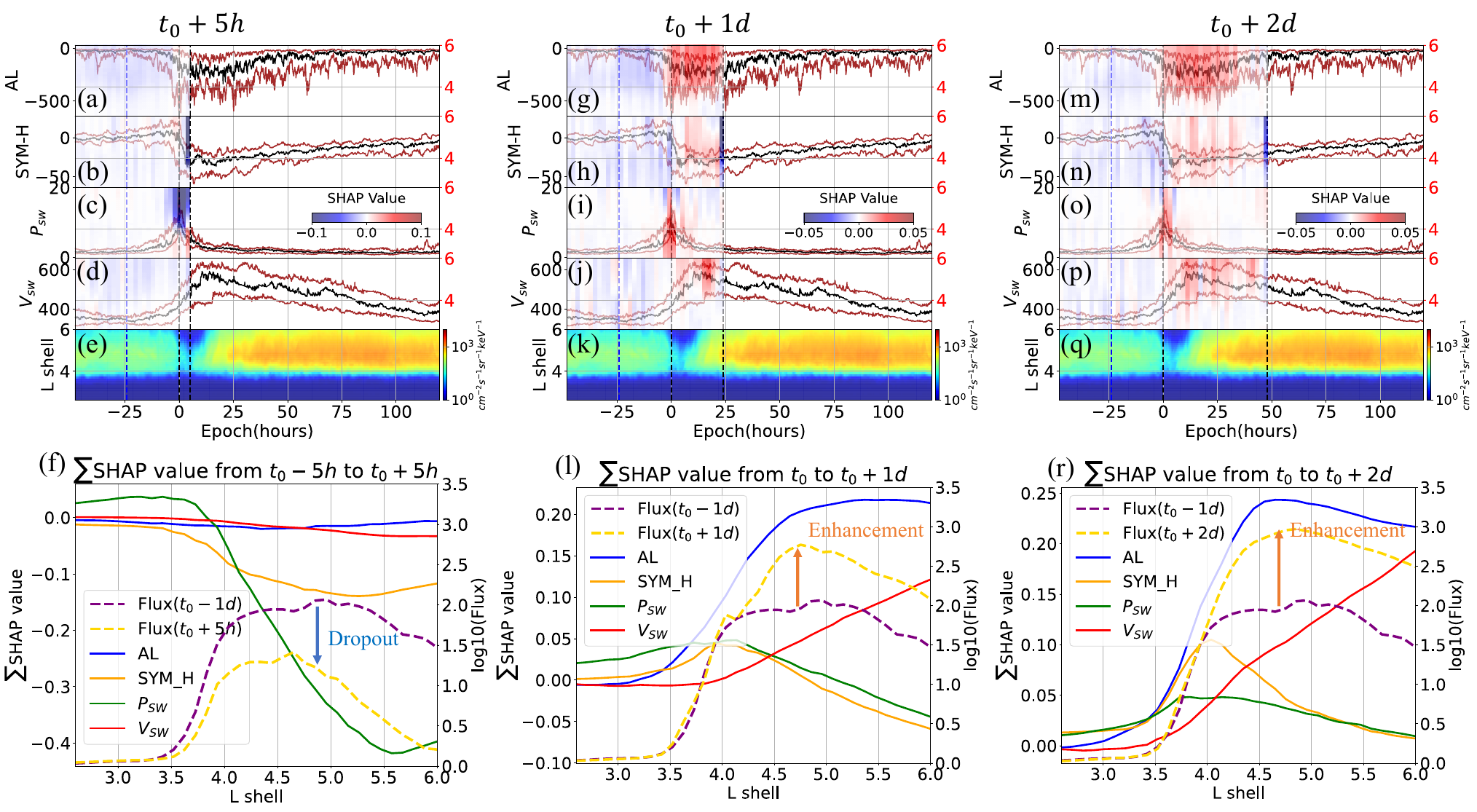}%
\caption{\label{Figure 2} Superposed epoch analysis and SHAP interpretation of geomagnetic indices and solar wind parameters for an enhancement event at three different times. The black lines in the upper three panels correspond to the median value, while the red lines correspond to the upper and lower quantiles. The vertical dashed line at $t_0 = 0$ is defined as the maximum of the solar wind pressure Psw. The vertical dashed line at $t=t_0-24h$(1d) is used to define the initial flux as used in panels (f, l, and r). The vertical dashed lines at $t_0+5h$ (a-f), $t_0+24h$ (g-l) and $t_0+48h$ (m-r) are the times used to evaluate the corresponding SHAP values. The median SHAP values of all events are then color-coded on the corresponding input features. (e, k and q) show the median flux results from machine learning model at 909 keV energy. The dashed lines in (f, l and r) show the initial flux (purple) and target flux (gold) at different L shells and times. The solid lines show the sum of median SHAP results of different indices: AL (blue), SYM-H (yellow), Psw (green) and Vsw (red) indicating their overall event importance.}%
\end{figure}

Figure 2 shows the results corresponding to the 19 enhancement events. Panels 2a to 2d show the time series of each of the inputs and the superposed results for AL, SYM-H, Psw, and Vsw displaying the median in black, and upper and lower quartiles in red. The median flux from the ML model in Figure 2e shows that the minimum flux level occurs roughly at $t_0 + 5h$ as indicated by the third vertical dashed line, following the dropout. The median SHAP results of the flux at $t_0 + 5h$ are colored-coded and superimposed on the input time series in the upper four panels of each column. It is clearly recognized that the negative contribution to the output flux comes mainly from the Psw maximum (Figure 2c) at around $t_0$, and SYM-H (Figure 2b) closer to the time of the observation at $t_0 + 5h$, where all other input values contribute to the output only weakly at this time. Figure 2f shows the sum of median SHAP values shown in Figures 2a-2d, of the different indices (solid lines), and the output flux result at $t_0 + 5h$ (yellow dashed line) compared to the initial flux at $t_0 - 1d$ (purple dashed line). The flux is seen to decrease at higher L-shells ($\mathrm{L} > 4$). The corresponding sum-SHAP results indicate that Psw (green line) is the primary contributor to the dropout as it has the lowest negative sum-SHAP values occurring at $\mathrm{L} > 4$. The feature attribution results here are consistent with previous studies during the storm time event \cite{turner2013storm}. They can be interpreted to mean that the dayside magnetopause causes trapped electrons to escape the outer boundary through magnetopause shadowing effect \cite{ukhorskiy2006storm} and occurs as the result of magnetopause moving abruptly inward in reaction to increased Psw. The enhanced SYM-H may be indicative of the Dst effect, inflating the outer radiation belt to larger L-shells and accelerating the rate of magnetopause shadowing loss.

Figures 2g to 2l show the result for the early acceleration phase at $t_0 + 1d$, when the enhancement of the fluxes starts to become apparent. The cluster of AL peaks (Figure 2g) occurring immediately after $t_0$ is brightly highlighted in red (i.e., positive SHAP values) which indicates that it contributes most to the acceleration of fluxes at higher L-shell ($\mathrm{L} > 4$). It is worth noting that the high solar wind speed (Figure 2j) also contributes to the acceleration at high L-shells, and we believe that is due to the collinearity between AL and Vsw, and will be discussed at the end of this section. Interestingly, there is some positive contribution to the flux from Psw (Figure 2i) at low L-shells (L$<$5) which may be associated with the compression occurring in the interior region of the ring current, but this conjecture requires further investigation.

Figures 2m to 2r show the SHAP results for the late acceleration phase, at $t_0 + 2d$, when the flux has reached its upper limit. As above, the results show that the highest contribution to the acceleration is from the cluster of AL peaks occurring immediately after $t_0$. Thus, it can be surmised that high-intensity continuous substorm activity produces enhanced fluxes of outer belt electrons through the injection of source and seed electrons \cite{jaynes2015source}, and continuous acceleration by enhanced chorus waves during such active times \cite{hua2022unraveling}.

\begin{figure}
\noindent\includegraphics[width=\textwidth]{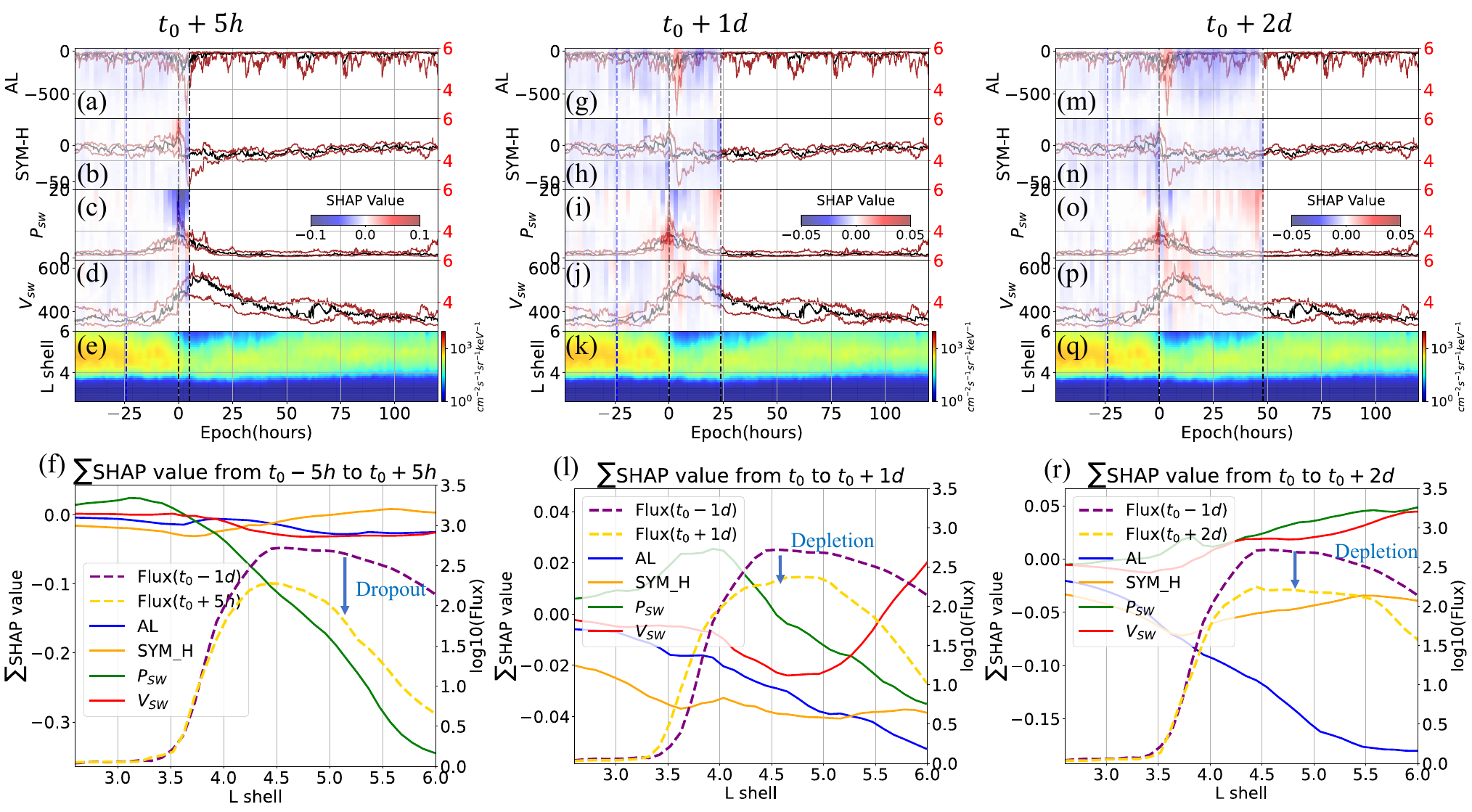}%
\caption{\label{Figure 3} Same as Figure 2 except for depletion events}%
\end{figure}

Figure 3 demonstrates the SHAP-enhanced superposed epoch analysis results corresponding to the 7 depletion events identified as described above. Figures 3a to 3f show the dropout process occurring in a very similar manner to the one in Figure 2e, with the minimum flux still occurring at around $t_0 + 5h$. The Psw enhancement is again seen to provide a negative contribution to the flux at higher L-shells ($\mathrm{L} > 4$). The results indicate that the effect of magnetopause shadowing is present in both groups of events, and electrons are quickly lost at higher L-shells because of the same process, with a smaller contribution coming from SYM-H in the present event.  Figures 3g to 3l show the early-stage development of the depletion of the fluxes and the SHAP results after pressure maximum at $t_0 + 1d$.  The late stage development of the depletion is displayed in Figures 3m-3r  for the period $t_0 + 2d$. Panel 3r shows the negative sum-SHAP values from AL and Figures 3g and 3m indicate that those negative AL contributions are the result of quiet substorm activities following the pressure enhancement. It is perhaps surprising at first that an extended period of low AL values act as the dominant contributor to the flux depletion, but this can be understood as follows: since the events in both acceleration and depletion groups experience a similar dropout process, the key determining feature of whether the flux ultimately becomes enhanced or depleted is essentially the total substorm activity that follows the dropout, as indicated by AL after $t_0$. The strong, continuous substorm activities (indicated by cluster of AL peaks) leads to subsequent overall acceleration, whereas quiet substorm activity (indicated by period of AL $\sim$ 0 nT) contributes to depletion. 
\begin{figure}
\noindent\includegraphics[width=\textwidth]{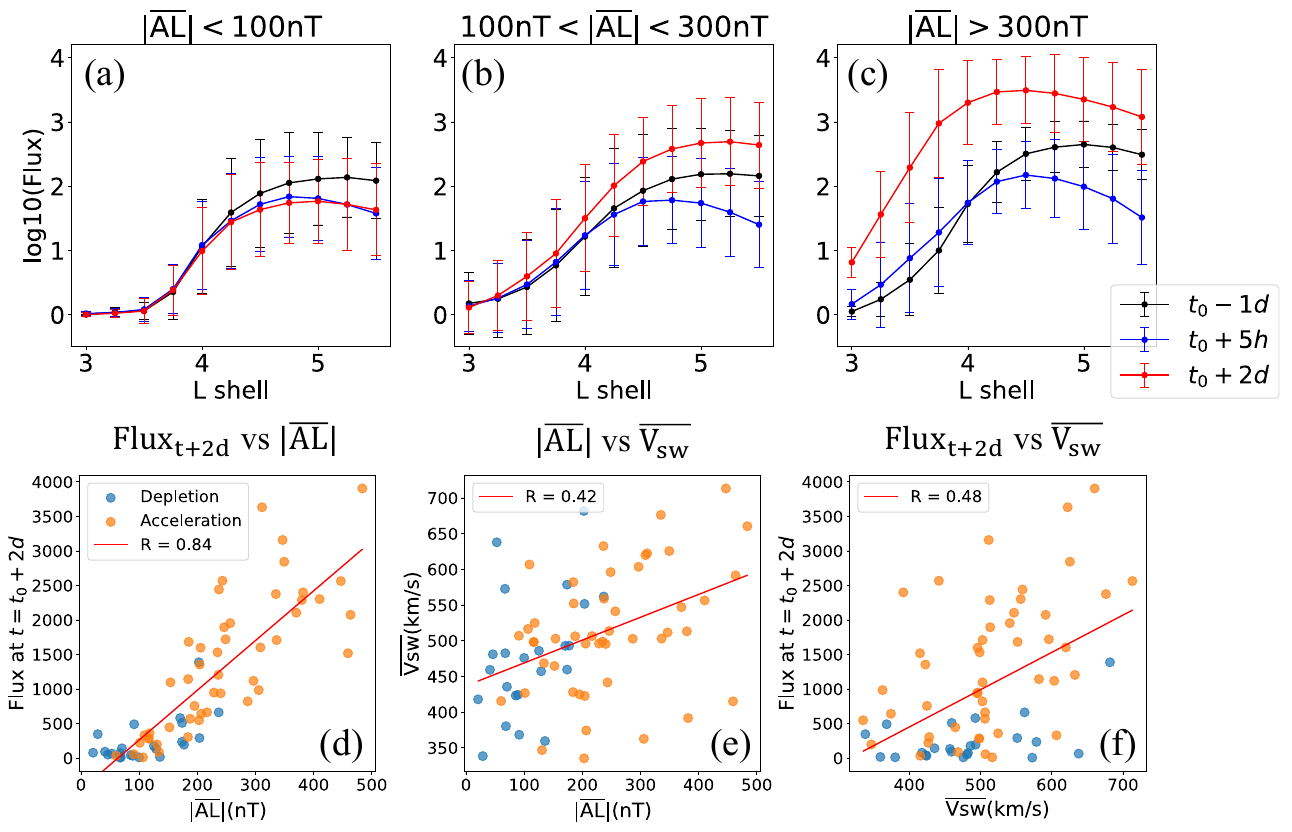}%
\caption{\label{Figure 4} The statistical results from 71 flux events and the relation to different geomagnetic indices and solar wind parameters. (a-c) The statistical flux measurements for 909 keV energy at $t_0 - 1d$ (black), $t_0 + 5h$ (blue) and $t_0+2d$ (red) with different $\overline{|\mathrm{AL}|}$, where the $\overline{|\mathrm{AL}|}$ is the average $|\mathrm{AL}|$ from $t_0$ to $t_0 + 2d$. The vertical error bars in each plot show the minimum and maximum range of the distribution of events in each L-shell bin. (d) shows the resulting flux at $t_0 + 2d$ from measurements at $\mathrm{L} = 5$ categorized by depletion (blue) and acceleration (orange) events, and its linear relation with $\overline{|\mathrm{AL}|}$. (e) shows the relation between $\overline{|\mathrm{AL}|}$ and $\overline{\mathrm{Vsw}}$, where $\overline{\mathrm{Vsw}}$ is average Vsw also from $t_0$ to $t_0 + 2d$. (f) shows the resulting flux at $t_0 + 2d$ and its relation to $\overline{\mathrm{Vsw}}$.}%
\end{figure}
To further confirm our conclusions obtained from the SHAP results, Figure 4 shows the statistical results from a larger dataset containing 71 events obtained in a similar way as previously described in Section 2. Figures 4a to 4c depict the average measured flux from Van Allen Probes at different L-shells and three distinct times. The epoch time $t_0$ still represents the maximum of solar wind dynamic pressure Psw, and the three times picked for comparison are the initial stages prior to the dropout ($t_0 - 1d$), the dropout stage ($t_0 + 5h$), and the final stage ($t_0 + 2d$). The average $|\mathrm{AL}|$ is calculated in the time interval ranging from $t_0$ to $t_0 + 2d$ as indicated on the SHAP results in Figures 2 and 3 for each event. Figure 4a demonstrates that under quiet substorm activity ($\overline{|\mathrm{AL}|} < 100 \mathrm{ nT}$), the flux first decreases for $\mathrm{L} > 4$ at the dropout stage, and the flux at the final stage remain similar to the flux level at the dropout stage. This indicates the relativistic electron flux cannot be accelerated in this range of substorm activity levels. Figure 4b shows a similar flux level to Figure 4a at both the initial and dropout stages but the flux at the final stage (red-colored curve) is enhanced to a level that is higher than the initial stage. 

Figure 4c presents the results corresponding to very strong substorm activity levels following the pressure maximum. The pre-storm flux is slightly higher than in Figures 4a and 4b, and the flux at the dropout stage is also higher at low L-shells. This could be related to the fact that these strong substorm events usually follow strong geomagnetic storms, which may affect the characteristics of the dropout (and will be examined in future studies). The fluxes at the final stage are seen to be enhanced to a significantly higher level than those corresponding to weak and moderate substorm conditions. 

Figure 4d then demonstrates the relation between the flux at $t_0 + 2d$ ($\mathrm{Flux}_{t+2d}$) at $\mathrm{L} = 5$ and the average absolute value of the AL index, $\overline{|\mathrm{AL}|} $. The results show a high correlation coefficient ($R = 0.84$) between the two values which is consistent with previous studies \cite{hua2022unraveling,mourenas2019impact}. Compared to prior studies, our interpretable ML method doesn't require extensive statistics based on different variables. Instead, it directly identifies the key influencing variable AL and the time ranges after Psw enhancement that are most significant. Our findings indicate that the depletion events can essentially be thought of as "non-acceleration" events, occurring when the substorms that follow the enhanced Psw are not sufficiently strong to enhance the flux above its prior level.

Although the SHAP profiles shown in Figures 2l and 2r suggest that the high solar wind speed contributes to the flux enhancement, it is not necessarily a condition that directly relates to the fluxes. This can be explained by noting that the training process of the ML model uses a feature selection method by adding the most informative drivers sequentially to the model \cite{ma2022modeling}, and shows that AL is the most important parameter, but adding Vsw does not affect model performance much. Figure 4c shows the correlation between the average AL and Vsw, and Figure 4f shows the relation between $\mathrm{Flux}_{t+2d}$ and the average Vsw. Neither AL nor the flux shows a good correlation to Vsw. In fact, the AL index can be modeled more accurately by the combination of solar wind speed and IMF Bz \cite{li2007prediction,mcpherron2015optimum}. The prolonged southward Bz together with high solar wind speed gives strong, continuous AL excursions, which drive the flux enhancement in the ML model. So the SHAP result of solar wind contribution may be due in large part to the collinearity between AL and Vsw.
\section{Conclusions and Discussion}
In this study, we investigated the response of relativistic radiation belt electron fluxes at 909 keV to various solar wind and geomagnetic disturbances. By combining an interpretable machine learning method and superposed epoch analysis, we reconstructed the fluxes and directly identified the key driving features of the electron flux enhancement and depletion events as a function of storm phase and L-shell.

The application of SHAP feature attribution method to the 19 enhancement and 7 depletion events has shown the following:
\begin{enumerate}
\item An increased solar wind dynamic pressure is the dominant contributor to the dropouts preceding both enhancement and depletion events. 
\item The high-intensity, continuous substorm activity following the pressure maximum (indicated by a cluster of AL peaks) contributes to the rapid increase of electron fluxes following the dropout during enhancement events.
\item  The quiet condition, or lack of substorm activity following the initial dropout, contributes to the decrease of electron flux during depletion events.
\end{enumerate}

To get more insight into our results, we performed a statistical study on 71 geospace disturbances events. The results show significant correlation between the resulting fluxes and average AL value following the solar wind pressure maximum. These results, in combination with our SHAP results, indicate that the depletion events can be thought of essentially as “non-acceleration” events that occur when substorm activity following the pressure maximum is not sufficient to accelerate the fluxes above its pre-storm level. 

Our study utilizes a novel approach to modeling and understanding the dynamics of Earth's radiation belt. Although the Superposed Epoch Analysis (SEA) is a popular method for identifying correlations between physical parameters involved in radiation belt dynamics, and hence inferring the causative driving factors, it has a number of important limitations. Specifically, the SEA cannot identify the roles of key parameters during rapid radiation belt flux changes, when the different parameters change simultaneously, or their roles in driving flux dynamics change as a function of time and/or L-shell; therefore, its results are often difficult to verify using physics-based models such as quasilinear simulations. In contrast, our interpretive machine learning model is not only capable of accurately reproducing the dynamics of the radiation belt, but it can also directly identify the key features corresponding to various significant dynamics, as they evolve in time and space, and which are shown to be in line with our physical understanding.

The results showing that average AL has a significant correlation with the resulting flux levels suggests that it is important to incorporate the AL index more directly into the radiation belt modeling. It is worth noting that the cluster of AL peaks can not only be used in identifying strong whistler-mode wave intensity \cite{li2009global}, but can also relate to the plasma frequency to gyrofrequency ratio ($\omega_{pe} / \omega_{ce}$) that affects the electron loss and energization efficiency and time scales \cite{agapitov2019time}. The purely data-driven results will serve as a baseline for future studies, that the density and wave models based on AL can be used in the radiation belt simulation. Furthermore, our conclusions should be applicable to a wide energy range of radiation belt electrons (e.g., 500 keV - 7 MeV), although there may be some quantitative differences for different energies. We only investigate a typical energy channel to demonstrate our method and leave others for future studies. The dropout occurrence rate and the magnitude of flux decrease during dropout may depend on the electron energy \cite{xiang2018statistical}. The acceleration of higher energy electrons also requires a longer time \cite{thorne2013rapid}, therefore affecting the SHAP values in the past times.

Finally, a number of caveats of our present work need to be mentioned. The uncertainty of the SHAP interpretability method comes primarily from the ML model itself, and there could be differences between the training and test datasets when explaining individual samples, so a model with very accurate performance should always be prioritized, as in the present case. The SHAP and other feature attribution methods usually assume feature independence, and this property requires us to choose the input features carefully, since it is clear that solar wind parameters ultimately control geomagnetic index values, albeit in complex ways. Although we chose our input features based on the strategy of adding the most informative predictors sequentially, there could still be hidden interactions that are ignored such as the solar wind and AL mentioned above. Possible solutions are using tree-like models that can have global interpretation and feature interaction \cite{lundberg2018consistent}, or using encoding and self-supervised methods \cite<e.g.>{he2020momentum} to map the input features to higher dimensions.

Despite the above caveats, we have demonstrated that a SHAP-enhanced superposed epoch analysis (SHESEA) has the unique ability to provide context for the standard SEA method, showing how independent variables control the dependent variable, how their roles vary as a function of time and how this behavior changes as a function of space.  This approach is general, and can be applied in a variety of situations where a standard SEA method is typically used, and is a novel way that ML can be used as an insight discovery tool in physical science.

\section{Open Research}

The event lists, data files and the detailed code to reproduce each figure are available at \url{https://doi.org/10.5281/zenodo.8145293} and the ORIENT machine learning model code example is available at \url{https://github.com/donglai96/ORIENT-M}.


\acknowledgments
The authors would like to thank the NASA SWO2R Award 80NSSC19K0239 and NASA CCMC award 80NSSC23K0324 for their generous support for this project (as well as subgrant 1559841 to the University of California, Los Angeles, from the University of Colorado Boulder under NASA Prime Grant agreement 80NSSC20K158). XC would like to thank Grant NASA ECIP 80NSSC19K0911. XC and QM would like to acknowledge the NASA Grant LWS 80NSSC20K0196. JB acknowledges support from the Defense Advanced Research Projects Agency under Department of the Interior award D19AC00009.


\bibliography{agusample.bib}

\end{document}